\documentstyle[12pt,amssymb]{article}

\def\fnote#1#2{\begingroup\def\thefootnote{#1}\footnote{#2}\addtocounter{footnote}{-1}\endgroup}

\def\inbar{\vrule height1.5ex width.4pt depth0pt}
\def\IB{\relax{\rm I\kern-.18em B}}
\def\IC{\relax\,\hbox{$\inbar\kern-.3em{\rm C}$}}
\def\ID{\relax{\rm I\kern-.18em D}}
\def\IE{\relax{\rm I\kern-.18em E}}
\def\IF{\relax{\rm I\kern-.18em F}}
\def\IG{\relax\,\hbox{$\inbar\kern-.3em{\rm G}$}}
\def\IH{\relax{\rm I\kern-.18em H}}
\def\II{\relax{\rm I\kern-.18em I}}
\def\IK{\relax{\rm I\kern-.18em K}}
\def\IL{\relax{\rm I\kern-.18em L}}
\def\IM{\relax{\rm I\kern-.18em M}}
\def\IN{\relax{\rm I\kern-.18em N}}
\def\IO{\relax\,\hbox{$\inbar\kern-.3em{\rm O}$}}
\def\IP{\relax{\rm I\kern-.18em P}}
\def\IQ{\relax\,\hbox{$\inbar\kern-.3em{\rm Q}$}}
\def\IR{\relax{\rm I\kern-.18em R}}
\def\IT{\relax{\rm I\kern-.18em T}}
\def\ZZ{\relax{\sf Z\kern-.4em Z}}

  \def\om{\omega}  \def\Om{\Omega} \def\si{\sigma}

 \def\cH{{\cal H}}  
   
\def\cO{{\cal O}} \def\cP{{\cal P}}

\def\pfrak{{\mathfrak p}}

\def\mathC{{\mathbb C}}   \def\mathF{{\mathbb F}}

\def\mathN{{\mathbb N}}   \def\mathQ{{\mathbb Q}}
   
\def\mathZ{{\mathbb Z}}

\def\bk{{\bar k}}

\def\fnote#1#2{\begingroup\def\thefootnote{#1}\footnote{#2}\addtocounter
{footnote}{-1}\endgroup}
\def\beq{\begin{equation}}
\def\eeq{\end{equation}}
\def\bea{\begin{eqnarray}}
\def\eea{\end{eqnarray}}
\def\llea#1{\label{#1}\eea}
\def\lleq#1{\label{#1}\eeq}
\let\nn=\nonumber
\def\tabroom{\hbox to0pt{\phantom{\Huge A}\hss}}
\def\notin{\ \hbox{{$\in$}\kern-.51em\hbox{/}}}

\def\lra{\longrightarrow}

\def\ra{{\rightarrow}}

\def\vphi{\varphi}

  \def\E1Fq{E_1/\IF_q}

  \def\rmH{{\rm H}}

   \def\rmdim{{\rm dim}}
   
\def\rmord{{\rm ord}}   \def\rmmod{{\rm mod}}
\def\rmsign{{\rm sign}} 
\def\rmtr{{\rm tr}}

\def\rmCo{{\rm Co}}   \def\rmEnd{{\rm End}}
   
\def\rmHW{{\rm HW}}   \def\rmIm{{\rm Im}}  
\def\rmSL{{\rm SL}}     \def\rmTr{{\rm Tr}}

\def\notdiv{{\relax{~|\kern-.35em /~}}}

  \def\E1Fq{E_1/\IF_q}

\begin{document}
\parindent=0pt

\hfill {\bf NSF$-$KITP$-$05$-$81}


\vskip 0.8truein

\centerline{\Large {\bf Arithmetic Spacetime Geometry from String
Theory }}

\vskip 0.6truein

\centerline{\sc Rolf Schimmrigk\fnote{\dagger}{email:
netahu@yahoo.com}}

\vskip .3truein

\centerline{\it Indiana University South Bend}
 \vskip .05truein
\centerline{\it 1700 Mishawaka Ave., South Bend, IN 46634}

\vskip 1truein

\baselineskip=19pt

\centerline{\bf ABSTRACT:} \begin{quote}
 An arithmetic framework to string compactification is described.
The approach is exemplified by formulating a strategy that allows
to construct geometric compactifications from exactly solvable
theories at $c=3$. It is shown that the conformal field theoretic
characters can be derived from the geometry of spacetime, and that
the geometry is uniquely determined by the two-dimensional field
theory on the world sheet. The modular forms that appear in these
constructions admit complex multiplication, and allow an
interpretation as generalized McKay-Thompson series associated to
the Mathieu and Conway groups. This leads to a string motivated
notion of arithmetic moonshine.
\end{quote}

\vfill

{\sc PACS Numbers and Keywords:} \hfill \break Math:  11G25
Varieties over finite fields; 11G40 L-functions; 14G10  Zeta
functions; 14G40 Arithmetic Varieties \hfill \break Phys: 11.25.-w
Fundamental strings; 11.25.Hf Conformal Field Theory; 11.25.Mj
Compactification

\renewcommand\thepage{}
\newpage
\pagenumbering{arabic}

\baselineskip=16pt
\parskip=2pt

\tableofcontents

\vfill \eject


\baselineskip=19pt
\parskip=.25truein
\parindent=0pt

\section{Introduction}

One of the long-standing questions in string theory has been the
problem to understand the nature of spacetime, in particular how
spacetime can emerge from a theory of 1-dimensional objects.
Phrased this way, the question is perhaps too vague to lend itself
to a direct approach, but it can be made more precise in the
context of exactly solvable string models\cite{g88, ks89}. In
these models the string input is described in terms of an
extension of the Virasoro algebra defined by affine Kac-Moody
algebras, and it is believed that such theories are closely
related to Calabi-Yau varieties.

It is the purpose of this paper to outline a strategy that allows
the construction of varieties from string theoretic objects, and
to show how this works in the simplest context of toroidal
compactifications, more precisely the class of Gepner models at
$c=3$. The main virtue of this type of string models is that their
structure is irreducible in a certain sense, in contrast to the
complicated motivic structure of higher dimensional varieties.
This can be expressed more concretely by noting that
cohomologically the only motivic invariant available is the
Jacobian of the curve. In higher dimensions the relation between
string theory and varieties is more complicated because this
irreducibility is lost. The strategy described here is general
enough to be applied in any dimension, but the rules that apply to
these different situations will be more involved.

The class of Gepner models with $c=3$ is a small set of three
models, simple enough to allow for a concise and coherent
discussion. Geometrically it concerns the elliptic curves defined
by Brieskorn-Pham polynomials
 \bea
 E_3 &=& \left\{(z_0:z_1:z_2) \in \IP_2~|~z_0^3+z_1^3+z_2^3=0
            \right\} \nn \\
 E_4 &=& \left\{(z_0:z_1:z_2) \in \IP_{(1,1,2)}~|~z_0^4+z_1^4+z_2^2=0
          \right\} \nn \\
 E_6 &=& \left\{(z_0:z_1:z_2) \in \IP_{(1,2,3)}~|~z_0^6+z_1^3+z_2^2=0 \right\},
 \eea
 which are expected to correspond to tensor models of $A_1^{(1)}$ theories at
various levels, equipped with the diagonal affine invariant.
Denote by $A_{1,k}^{(1)}$ the affine Lie algebra $A_1^{(1)}$ at
affine level $k$. It is conjectured that there exist relations
 \beq
 E_i ~\sim ~ \left(\bigotimes_{k_{ij}}
          A_{1,k_{i,j}}^{(1)}\right)_{\rm GSO},
 \eeq
 where the subscript GSO indicates the
supersymmetry projection, the affine levels are given by
 \beq
 (k_{i,1},k_{i,2},k_{i,3}) \in \{(1,1,1), (2,2,0), (4,1,0)\},
 \eeq
and 0 denotes a trivial factor.

The first part of this paper completes the modular analysis of the
class of elliptic Brieskorn-Pham models, initiated in \cite{su03},
and continued as part of a discussion with a different focus in
\cite{ls04}. More precisely, the following string theoretic
interpretation of the L-function of the degree six elliptic
Brieskorn-Pham curve is shown to hold. Denote by $\Gamma_0(N)$ the
congruence subgroup of $\rmSL(2,\mathZ)$ defined by matrices that
are upper triangular mod some integer $N$, \beq \Gamma_0(N) =
\left\{\left(\matrix{a&b\cr c&d\cr}\right) \in \rmSL_2(\ZZ) ~{\Big
|}~\left(\matrix{a&b\cr c&d\cr}\right) \sim \left(\matrix{*&*\cr
0&*\cr}\right) ~(\rmmod~N)\right\}, \eeq and denote by \beq
\eta(q) = q^{1/24}\prod_{n=1}^{\infty} (1-q^n) \eeq the Dedekind
eta function, where $q=e^{2\pi i \tau}$. Let furthermore
$c^k_{\ell,m}(\tau)$ be the Kac-Peterson string functions of
$A_{1,k}^{(1)}$, defined in Section 3, and denote by $f\otimes
\chi$ the twisted modular form defined as
 \beq
  f(q) \otimes \chi = \sum_{n} a_n \chi(n) q^n.
 \lleq{twistedform}

{\bf Theorem 1.}~{\it The Mellin transform of the elliptic curve
$E_6$ is a modular form in $S_2(\Gamma_0(144))$ which is
determined by the Hecke indefinite modular form
$\Theta^1_{1,1}(\tau)= \eta^3(\tau) c^1_{1,1}(\tau)$ of the string
function $c^1_{1,1}(\tau)$ at conformal level $k=1$ as $
f_{\rmHW}(E_6,q) = \Theta^1_{1,1}(q^6)^2\otimes \chi_3$,  where
$\chi_3(p)=\left(\frac{3}{p}\right)$ is the Legendre symbol.}

Combining this result with the results of \cite{su03} and
\cite{ls04} leads to the following complete identification of the
elliptic Brieskorn-Pham curve L-functions in terms of string
theoretic modular forms.

{\bf Theorem 2.}~{\it The Mellin transforms $f_{\rmHW}(E_i,q)$ of
the Hasse-Weil L-functions $L_{\rmHW}(E_i,s)$ of the curves $E_i,
i=3,4,6$ are modular forms $f_{\rmHW}(E_i,q) \in
S_2(\Gamma_0(N))$, with $N\in \{27, 64, 144\}$ respectively. These
forms factor into products of Hecke indefinite modular forms
$\Theta^k_{\ell,m}(\tau) = \eta^3(\tau)c^k_{\ell,m}(\tau)$,
determined by the string functions $c^k_{\ell,m}(\tau)$, as
follows}
 \bea
f_{\rmHW}(E_3,q) &=& \Theta^1_{1,1}(q^3)\Theta^1_{1,1}(q^9) \nn \\
f_{\rmHW}(E_4,q) &=& \Theta^2_{1,1}(q^4)^2\otimes \chi_2 \nn \\
f_{\rmHW}(E_6,q) &=& \Theta^1_{1,1}(q^6)^2\otimes \chi_3,
 \llea{hw-as-thetas}
{\it where $\chi_n(p)=\left(\frac{n}{p}\right)$ denotes the
Legendre symbol.}

The appearance of the Legendre symbols $\chi_n$ in the above
theorems can be explained physically. Their origin can be found in
the algebraic number field that is determined by the conformal
weights of the underlying conformal field theory via the Rogers
dilogarithm.

{\bf Corollary.}~{\it The twist characters $\chi_n$ are the
quadratic characters of the fields of quantum dimensions of the
underlying affine Lie algebra $A_1^{(1)}$.}

More concretely, these quadratic characters determine the
factorization behavior of the rational primes in the field
extensions determined by quantum dimensions of the conformal field
theory. In the discussion of the cubic curve $E_3$ in \cite{su03}
no such character appeared because the field of quantum dimensions
in that case does not define an extension of the field of rational
numbers, but is the field $\mathQ$ itself.

Theorem 2 shows that certain conformal field theoretic characters
that enter the string partition functions of these models are
determined by the arithmetic properties of the topological part of
spacetime, and vice versa. Mathematically it establishes a
relation between the arithmetic of elliptic curves and the theory
of affine Kac-Moody algebras.

In the second part of this paper it is shown that it is possible
to reverse the above logic and to formulate a set of assumptions
which uniquely determine the conformal field theoretic modular
forms of the string which appear in Theorem 2, without any a
priori notions from geometry. Given this identification of
potentially geometric forms derived from the string, one can ask
whether it is possible to construct the geometry of spacetime
purely from string theoretic ingredients. In the context of
abelian varieties this can be achieved by the theory initiated by
Klein, Hurwitz and Hecke, and further developed by Eichler and
Shimura.

The relation described here, between the arithmetic geometry of
elliptic curves on the one hand, and the theory of affine Lie
algebras on the other, can also be used to establish a new link
between Kac-Moody algebras and modular moonshine for some finite
sporadic groups, such as the Mathieu group and the Conway group.
The theory of generalized Kac-Moody algebras has been used by
Borcherds \cite{b92} to prove the monstrous moonshine conjecture
of Conway and Norton \cite{cn79}, providing an interpretation of
hauptmoduln as McKay-Thompson series. The idea to interpret
modular forms of higher weight in a similar way has been discussed
by Mason \cite{m90} and Martin \cite{m96} for the largest Mathieu
group $M_{24}$ as well as the Conway group $\rmCo$. Combining
their work with the results described here leads to an
interpretation of string theoretic modular forms as generalized
McKay-Thompson series in the context of affine Lie algebras.

The outline of the paper is as follows. In Section 2 the
Hasse-Weil L-function is computed and the geometries are analyzed
to the extent necessary to compute the geometric conductors of the
curves. These conductors determine the levels of the modular forms
expected from the modularity theorem of Wiles, Taylor, and Breuil
et.al. In Section 3 the basic building blocks of the affine
Kac-Moody algebra are determined at the levels $k=1,2,4$. Section
4 completes the proof of the string theoretic interpretation of
the Hasse-Weil L-functions for the Brieskorn-Pham type elliptic
curves. In Section 5 a set of criteria is formulated which
geometric modular forms should satisfy. These constraints uniquely
determine the Hasse-Weil modular forms of the elliptic
Brieskorn-Pham curves considered in this paper. Section 6
describes how the arithmetic geometry of these curves can be
recovered from the modular forms. In Section 7 it is shown that
the modular forms encountered here admit complex multiplication in
the sense of Shimura-Taniyama, in agreement with the conjecture of
\cite{lss04}. Section 8 points out that the Hasse-Weil modular
forms can be interpreted as generalized McKay-Thompson series of
either the largest Mathieu group $M_{24}$, or the Conway group
$\rmCo$ determined as the automorphism group of the Leech lattice,
thereby leading to sporadic modular moonshine.

\vskip .2truein

\section{From geometry to modularity}

\subsection{Hasse-Weil L-functions}

The strategy in this paper is to use the arithmetic of elliptic
curves to produce modular forms and to show that these modular
forms admit a conformal field theoretic interpretation. More
details, physical motivation, historical remarks, and further
references can be found in \cite{su03}. The idea is to use the
congruent zeta function of Artin, Schmidt, and Weil
 \beq
 Z(X/\IF_p,t) = \exp\left(\sum_{r\geq 1} \frac{N_{r,p}(X)}{r}t^r
\right),
 \eeq
 where $N_{r,p}(X)$ denotes the number of solutions of
the variety $X$ over the finite field extension $\mathF_{p^r}$ and
$t$ is a formal variable. Schmidt has shown that the zeta function
of curves $C$ of genus $g$ factorizes into the rational form
 \beq
 Z(C,t) = \frac{\cP^{(p)}(t)}{(1-t)(1-pt)},
 \eeq
 where $\cP^{(p)}(t)$ is a polynomial of degree $2g$. It is this
polynomial which is used to define the Hasse-Weil L-function. The
Hasse-Weil L-function globalizes the local congruent functions by
combining the information at all primes
 \beq
 L_{\rm HW}(C,s) =
 \prod_{p ~{\rm prime}} \frac{1}{\cP^{(p)}(p^{-s})}.
 \eeq

For elliptic curves the polynomials $\cP^{(p)}(t)$ can be written
as
 \beq
  \cP^{(p)}(t) = 1+\beta_1(p)t + \delta(p)pt^2,
  \eeq
  where $\beta_1(p)$ and $\delta(p)$ depend on the properties of the
reduction over the curve of the finite field $\IF_p$. At the good
primes $\beta_1(p)$ is given by
 \beq
 \beta_1(p) = N_{1,p}(E) - (p+1),
 \eeq
 while at the bad primes its structure depends on the
type of the singularity
 \beq
 \beta_1(p) = \left\{\begin{tabular}{c l} $\pm 1$ &if the singularity at $p$
is a node  \\
   0  &if the singularity at $p$ is a cusp \tabroom \\
         \end{tabular}\right\}.
   \eeq
Here the sign in the first case depends on whether the node is
split or non-split. $\delta(p)$ finally is given as
 \beq
 \delta(p) = \left\{\begin{tabular}{c l} 0 &if $p$ is a bad prime \\
1  &if $p$ is a good prime \tabroom \\
\end{tabular}\right\}.
 \eeq
 With these ingredients the Hasse-Weil L-function can then be defined
as
 \beq
 L_{\rm HW}(X,s) =
 \prod_{p\in S} \frac{1}{1+\beta_1(p)p^{-s}} \prod_{p \notin S}
\frac{1}{1+\beta_1(p)p^{-s}+p^{2-2s}},
 \eeq
 where $S$ denotes the set of bad primes.

The Hasse-Weil L-functions of $E_3$ and $E_4$ can be found in
\cite{su03} and \cite{ls04} respectively. Here the analysis is
completed by computing the L-functions for the degree six elliptic
curve. Table 1 contains the cardinalities and the resulting
coefficients $\beta_1(p)$ for the lower primes.
\begin{center}
\begin{tabular}{c| c c c  c  c c c c c c c c c c}

Prime $p$     &2   &3   &5  &7  &11 &13  &17  &19  &23  &29  &31
              &37  &41  &43                 \tabroom \\
\hline

$N_{1,p}$     &3   &4   &6  &4  &12 &12  &18  &28  &24 &30 &28
              &48  &42  &52  \tabroom \\

$\beta_1(p)$  &0   &0   &0  &$-4$ &0 &$-2$ &0 &8   &0   &0 &$-4$
              &10  &0   &8     \tabroom \\
\hline
\end{tabular}
\end{center}

\centerline{{\bf Table 1.}~{\it Cardinalities $N_{1,p}(E_6)$ for the
elliptic curve of degree six $E_6 \subset \IP_{(1,2,3)}$.}}

By taking into account the fact that the bad primes for $E_6$ are
given by $p=2,3$ one finds that its L-function takes the form \beq
L_{\rmHW}(E_6,s) = 1
 + \frac{4}{7^s} + \frac{2}{13^s}  - \frac{8}{19^s} - \frac{5}{25^s}
 + \frac{4}{31^s} + \cdots \eeq

To any $L-$series $L=\sum_n a_n n^{-s}$ one can associate a
$q-$series by replacing $n^{-s} \mapsto q^n$. Doing so leads to
what might be called a Hasse-Weil $q-$expansion. Table 2
summarizes the results of the expansions for the curves discussed
here and in \cite{su03, ls04}.

\begin{center}
\begin{tabular}{c| l}
Curve  &Hasse-Weil $q-$expansion \tabroom \\
\hline

$E_3$  &$ q - 2q^4 - q^7 + 5q^{13} + 4q^{16} - 7q^{19} - 5q^{25} +
2q^{28} - 4q^{31} + 11q^{37} + \cdots$
\tabroom \\

$E_4$  &$  q +2q^5 -3q^9 -6q^{13} +2q^{17} -q^{25} +10q^{29}
+2q^{37} + 10q^{41} + \cdots $ \tabroom \\

$E_6$  &$q + 4q^7 + 2q^{13}  - 8q^{19} - 5q^{25} + 4q^{31} -
10q^{37} - 8q^{43} + \cdots$ \tabroom \\
 \hline
\end{tabular}
\end{center}

\centerline{{\bf Table 2.}{\it ~The Hasse-Weil $q-$expansions
$f_{\rmHW}(E_i,q)$ for the curves $E_3,E_4,E_6$.}}

\subsection{Modular forms from the Hasse-Weil L-function}

The proof of the Shimura-Taniyama conjecture in the modular curve
theorem says that the $q-$expansion associated to $L_{\rmHW}(E,s)$
is a modular form of weight two and some modular level $N$. More
concretely, the weight of a modular form which is a Hecke
eigenfunction can be determined from its multiplicative structure,
while the level can be obtained from the conductor of the curve
and checked via the Fricke operator. Both of these ingredients are
provided by the Hecke algebra of operators that act on modular
forms. The basic ingredients in the present context can be
summarized as follows.

One possible way to define Hecke operators $T_n^w$, $n \in
\mathN$, on modular forms $f(\tau)$ of weight $w$ with character
$\chi$ is by considering the following linear combination of
shifts
 \beq
   T_n^wf(\tau)= \frac{1}{n} \sum_{ad=n} \chi(a)a^w \sum_{0\leq b<d}
        f\left(\frac{a\tau + b}{d}\right).
 \eeq
 If $f(q)=\sum_{n=0}^{\infty} a_nq^n$ is the $q-$expansion of the
 modular form, the expansion image $T_n^wf(q) =
 \sum_{n=0}^{\infty}b_mq^m$ is given by
 \beq
  b_m = \sum_{d|(m,n)} \chi(d)d^{w-1}a_{(mn/d^2)}.
 \eeq
 The operators $T_n^w$ map the space of cusp forms
 $S_w(\Gamma_0(N),\chi)$ into itself, and for prime $p$ the
 operator simplifies on cusp forms to
  \beq
   T_p^wf(q) = \sum_{n=1}^{\infty} a_{np}q^n + \chi(p)p^{w-1}
   \sum_{n=1}^{\infty} a_nq^{np}.
  \eeq
Then
   \bea
  T^w_{mn} &=& T^w_mT^w_n,~~~~~~m,n~{\rm coprime}, \nn \\
  T^w_{p^{e+1}} &=& T^w_{p^e}T^w_p - p^{w-1}T^w_{p^{e-1}}.
  \eea

 A special operator has
to be considered at primes $p$ which divide the level $N$ of the
form. This is the Atkin-Lehner operator \cite{al70},
 often denoted by $U_k(p)$
 \beq
 U_pf(q) = \sum a_{pn}q^n.
 \eeq
 For eigenforms of these operators the operator
structure translates into identical relation between their
coefficients $a_n$
 \bea
 a_{mn}&=& a_ma_n~~~(m,n)=1 \nn \\
 a_{p^{n+1}} &=& a_{p^n}a_p - p^{k-1}a_{p^{n-1}},~~~~
                {\rm for}~p\notdiv N \nn \\
 a_{p^n} &=& (a_p)^n,~~~~{\rm for}~p|N.
 \eea

 The Hasse-Weil forms $f_{\rmHW}(E_i,q)$ of the elliptic curves of
 Brieskorn-Pham type satisfy these relations with $w=2$, hence define normalized
 cusp Hecke eigenforms of weight two.

This leaves the question what the level $N$ is of the modular
forms listed in Table 2. The answer to this problem involves two
ingredients, Weil's conductor conjecture, and an involution $w_N$
that preserves the space of cusp forms on $\Gamma_0(N)$ and is
defined by
 \beq
 w_N(f)(\tau) = N\tau^2f\left( {\small
\frac{-1}{N\tau}}\right).
 \eeq
  Weil observed that the geometric
conductor of an elliptic curve is in fact the level of the modular
form \cite{w67}, hence it provides guidance which can then be used
for the Fricke involution $w_N$.

The first step therefore is to compute the conductors of the
elliptic curves $E_i$. For an elliptic curve this is a quantity
which is determined both by the rational primes for which the
reduced curve degenerates, i.e. the bad primes, as well as the
degeneration type. The conductor of the curve can be computed by
first transforming the Fermat cubic into a Weierstrass form and
then applying Tate's algorithm \cite{t75}. Since it is necessary
to consider fields of characteristic 2 and 3, the usual (small)
Weierstrass form $y^2=4x^3+Ax+B$ is not appropriate. Instead one
has to consider the generalized Weierstrass form given by \beq
E:~~~y^2 + a_1xy +a_3y =x^3 + a_2x^2 + a_4x +a_6,
\lleq{genweierstrass} where the unusual index structure indicates
the weight of the coefficients under admissible transformations
which preserve this form \beq (x,y)
~\mapsto~(u^2x+r,u^3y+u^2sx+t),\eeq with $r,s,t\in K$ and $u\in
K^{\times}$ if $E$ is defined over the field $K$. Curves of this
type can acquire certain types of singularities when reduced over
finite prime fields $\IF_p$. The quantity which detects such
singularities is the discriminant
 \beq
 \Delta = \frac{c_4^3-c_6^2}{1728},
 \eeq
 where
 \bea
 c_4 &=& b_2^2 - 24b_4 \nn  \\
 c_6 &=& - b_2^2 + 36 b_2 b_4 - 216b_6,
 \eea
 with
 \bea
b_2 &=& a_1^2+4a_2 \nn \\ b_4&=&a_1a_3 + 2a_4 \nn \\
b_6 &=& a_3^2+4a_6.
 \eea
 The curve $E$ acquires singularities at
those primes $p$ for which $p|\Delta$. The singularity types that
can appear have been classified by Kodaira and N\'eron \cite{n64},
and are indicated by Kodaira's symbols. ${\rm I}_0$ describes the
smooth case, ${\rm I}_n, (n>0)$ involve bad multiplicative
reduction, and ${\rm I}_n^*, {\rm II}, {\rm III}, {\rm IV}, {\rm
II}^*, {\rm III}^*, {\rm IV}^*$ denote bad additive reduction.

The conductor itself depends on the detailed structure of the bad
fiber and the discriminant. Conceptually, it is defined as an
integral ideal of the field $K$ over which the elliptic curve $E$
is defined. By a result of Ogg \cite{o67} this ideal is determined
by the number $s_p$ of irreducible components of the singular
fiber at $p$ as well as the order $\rmord_p \Delta_{E/K}$ of the
discriminant $\Delta_{E/K}$ at $p$. In the present cases the
curves are defined over the field $K=\mathbb{Q}$, hence the ring
of integers is a principal domain. The conductors can therefore be
viewed as numbers defined by \beq N_{E/\mathbb{Q}} = \prod_{{\rm
bad}~p} p^{f_p},\eeq where the exponent $f_p$ is given by \beq f_p
= \rmord_p \Delta_{E/\mathbb{Q}} + 1 -s_p.\eeq

$E_3$: The Fermat cubic can be transformed into the form
$v^2-9v=u^3-27$,
 which in turn can be transformed
further by completing the square and introducing the variables
$x=u$ and $y=v-5$, leading to the affine curve \beq y^2 + y = x^3
- 7. \eeq This curve has discriminant $\Delta(E_3) =-3^9$ and
$j-$invariant $j=0$, while the singular fiber resulting from
Tate's algorithm is of Kodaira type IV$^*$ with $s_3=7$
components. Using the fact that $\rmord_3(\Delta(E_3)) =9$ leads
to the conductor $N=27$.

$E_4$: The Brieskorn-Pham quartic can be transformed into the
Weierstrass form \beq v^2=u^3+u, \eeq which has discriminant
$\Delta(E_4) = -2^6$. It follows that $\rmord_2(\Delta(E_4)) = 6$,
and since the singular fiber resulting from Tate's algorithm is of
Kodaira type II with $s_2=1$ component, this leads to the
conductor $N=64$.

$E_6$: The Weierstrass form of this curve is given by \beq v^2 =
u^3 - 1.\eeq The discriminant of this curve is
 $\Delta(E_6) = - 2^4\cdot 3^3$, hence both 2 and 3 are bad primes,
 as expected, and
we have $\rmord_2(\Delta(E_6))=4$ and $\rmord_3(\Delta(E_6))= 3$.
The singularity types are given by the fibers of the type \beq
p=2:~~~{\rm II},~~~~p=3:~~~ {\rm III}.\eeq Hence Ogg's result
leads to the conductor $N= 144$.

\subsection{Dimension of spaces of modular forms}

The conductor computations above show that the modular forms
derived from the Hasse-Weil L-functions considered here are cusp
forms of conductors
 \beq
 N\in \{27,64,144\}.
 \eeq
 It is useful to
compute the dimensions of the corresponding spaces
$S_2(\Gamma_0(N))$. The dimension of the space of cusp forms at
fixed level $N$ can be found in many places, e.g. \cite{gs71a}. It
turns out to be given by the genus of an algebraic curve
$X(\Gamma_0(N))$ constructed from the congruence group
$\Gamma_0(N)$, for reasons that will become clear further below.

{\bf Proposition.} {\it The dimension of the space
$S_2(\Gamma_0(N))$ is given by}
 \beq
 \rmdim~ S_2(\Gamma_0(N)) = g(X(\Gamma_0(N))),
 \eeq
 {\it where the genus of $X(\Gamma_0(N))$ is determined by}
 \beq
  g(X(\Gamma_0(N)) = 1 + \frac{\mu(N)}{12} -
 \frac{\nu_2(N)}{4} - \frac{\nu_3(N)}{3} -
\frac{\nu_{\infty}(N)}{2},
 \lleq{modulargenus}
 {\it where $\mu(N), \nu_2(N),\nu_3(N), \nu_{\infty}(N)$ denote the index of
$\Gamma_0(N)$, the number of elliptic points of order 2,3, and the
number of $\Gamma_0(N)-$inequivalent cusps respectively.}

The ingredients of this dimension result only depend on the
modular level $N$ and its prime divisors, and closed formulae are
available.

{\bf Proposition.}
 {\it The index $\mu(N)$ of $\Gamma_0(N)$ is given by}
  \beq
  \mu(N) = N\prod_{p|N}\left(1+\frac{1}{p}\right),
  \eeq
 {\it where the product is over all prime divisors of $N$.
  $\Gamma_0(N)$ has elliptic elements (i.e. $\gamma$ with
$|\rmTr(\gamma)| < 2$). Their numbers are given by } \bea
 \nu_2(N) &=& \left\{\begin{tabular}{l l}
     0 &if $4|N$  \\
    $\prod_{p|N}\left(1+\left(\frac{-4}{p}\right)\right)$  &otherwise
             \tabroom \\
     \end{tabular} \right\} \nn \\
 \nu_3(N)
  &=& \left\{\begin{tabular}{l l}
    0 &if $2|N$ or $9|N$   \\
$\prod_{p|N}\left(1+\left(\frac{-3}{p}\right)\right)$  &otherwise
\tabroom \\
\end{tabular} \right\},
\eea {\it where $\left(\frac{\cdot}{p}\right)$ denotes the
Legendre symbol. Finally, the number of cusps is given by}
 \beq
 \nu_{\infty}(N) = \sum_{0<d|N} \phi((d,N/d)),
 \eeq
 {\it where
$(d,N/d)$ denotes the greatest common divisor and $\phi(n)$ is the
Euler totient function.}

The results for $N=27$ and $N=64$ have been described e.g. in
\cite{su03} and \cite{ls04}. For the space $S_2(\Gamma_0(144))$
one obtains $\mu(144) = 288$. There are no elliptic points of
order 2, $\nu_2(144) = 0$, because 144 is divisible by 4. There
are also no elliptic points of order 3, because $9|144$, hence
$\nu_3(144) = 0$. Finally, the number of $\Gamma_0(144)$
independent cusps is $\nu_{\infty}(144) = 24$. Plugging these
ingredients into the genus formula gives $g(X(\Gamma_0(144))) =
13$. Collecting everything one finds that the curves $E_i$ lead to
the results collected in Table 3.

\begin{center}
\begin{tabular}{c| c c c c c}

Curve  &$\Delta$       &Type         &$s_p$  &$N$
       &$\rmdim~S_2(\Gamma_0(N))$ \tabroom \\
\hline

$E_3$  &$-3^9$      &IV$^*$  &7      &27   &1  \tabroom \\

$E_4$  &$-2^6$     &II      &1      &64   &3  \tabroom \\

$E_6$  &$-2^4\cdot 3^4$
                   &II, III &(1,2)  &144  &13  \tabroom \\

\hline
\end{tabular}
\end{center}

\centerline{{\bf Table 3.}~{\it Relevant characteristics for the
elliptic curves $E_i$.}}

\section{Affine Kac-Moody Algebras}

\subsection{$N=2$ supersymmetric models}

Supersymmetric string models can be constructed in terms of
conformal field theories with N=2 supersymmetry. The simplest
class of N=2 supersymmetric exactly solvable theories is built in
terms of the affine SU(2)$_k$ algebra at level $k$ as a coset
model \beq \frac{{\rm SU(2)}_k \otimes {\rm U(1)}_2}{{\rm
U(1)}_{k+2,{\rm diag}}}. \eeq Coset theories $G/H$ lead to central
charges of the form $c_G - c_H$, hence the supersymmetric affine
theory at level $k$  still has central charge $c_k=3k/(k+2)$. The
spectrum of conformal weights $\Delta^{\ell}_{q,s}$ and
U(1)$-$charges $Q^{\ell}$ of the primary fields
$\Phi^{\ell}_{q,s}$ at level $k$ is given by
\begin{eqnarray} \Delta^{\ell}_{q,s} &=&
\frac{\ell (\ell +2)-q^2}{4(k+2)} + \frac{s^2}{8} \nonumber \\
Q^{\ell}_{q,s} &=& - \frac{q}{k+2} + \frac{s}{2}, \end{eqnarray}
where $\ell\in \{0,1,\dots,k\}$, $\ell+q+s \in 2\mathbb{Z}$, and
$|q-s|\leq \ell$. Associated to the primary fields are characters
defined as
\begin{equation}
\chi^k_{\ell,q,s}(\tau, z,u) = e^{-2\pi i u} {\rm
tr}_{\cH^{\ell}_{q,s}} e^{2\pi i\tau (L_0 -\frac{c}{24})} e^{2\pi
i J_0}, \end{equation} where the trace is to be taken over a
projection $\cH^{\ell}_{q,s}$ to a definite fermion number (mod 2)
of a highest weight representation of the (right-moving) $N=2$
algebra with highest weight vector determined by the primary
field. It is of advantage to express these maps in terms of the
string functions and theta functions, leading to the form
\begin{equation} \chi^k_{\ell,q,s}(\tau, z, u) = \sum
c^k_{\ell,q+4j-s}(\tau) \theta_{2q+(4j-s)(k+2),2k(k+2)}(\tau, z,u)
\end{equation} because it follows from this representation that
the modular behavior of the $N=2$ characters decomposes into a
product of the affine SU(2) structure in the $\ell$ index and into
$\Theta$-function behavior in the charge and sector index. The
string functions $c^k_{\ell,m}$ are given by \beq c^k_{\ell, m}
(\tau) = \frac{1}{\eta^3(\tau)}
\sum_{\stackrel{\stackrel{-|x|<y\leq |x|}{(x,y)~{\rm
or}~(\frac{1}{2}-x,\frac{1}{2}+y)}}{\in
\mathZ^2+\left(\frac{\ell+1}{2(k+2)},\frac{m}{2k}\right)}}
\rmsign(x) e^{2\pi i \tau((k+2)x^2-ky^2)} \lleq{stringfctn} while
the classical theta functions $\theta_{m,k}$ are defined as
 \beq
 \theta_{n,m}(\tau,z,u) = e^{-2\pi i m u} \sum_{\ell \in \mathZ +
 \frac{n}{2m}} e^{2\pi i m \ell^2 \tau + 2\pi i \ell z}.
 \eeq
 It
follows from the coset construction that the essential ingredient
in the conformal field theory is the SU(2) affine theory.

\subsection{Hecke indefinite modular forms}

This section records the Hecke indefinite forms
$\Theta^k_{\ell,m}(\tau)= \eta^3(\tau)c^k_{\ell,m}$ that are
associated to the independent string functions at the level
$k=1,2,4$. The number of string functions defined in
(\ref{stringfctn}) at level $k$ is restricted by the level $k$ via
the constraints \beq 0\leq \ell \leq k,~~~~-\ell \leq m \leq
2k-\ell,~~~~\ell = m ~\rmmod ~2. \eeq Not all the resulting string
functions are independent; there are relations which can be
encoded as follows \cite{g88} \beq c^k_{\ell,m} = c^k_{\ell,-m} =
c^k_{\ell,m+2k} = c^k_{k-\ell,m+k}. \eeq These identifications
lead to a unique modular form at level 1, three independent forms
at level 2, and seven independent forms at level 4. The expansions
of these theta forms are collected in Table 3. These are important
in particular for the inverse problem of finding 'geometric' forms
at their various levels, considered in the second part of this
paper.

\begin{center}
\begin{tabular}{c| l }

Level $k$ &Hecke modular form $\Theta^{\ell}_m$ \tabroom \\ \hline

1 &$\Theta^1_{1,1}(\tau)
     = q^{1/12}(1 - 2q - q^2 + 2q^3 + q^4 + 2q^5 - 2q^6 - 2q^8 + \cdots)$
      \tabroom \\

\hline

2 &$\Theta^2_{0,0}(\tau)
    = q^{1/16}(1 - 2q + q^3 + q^5 - 2q^6 + 2q^7 - 2q^{12} + \cdots)$
     \tabroom \\

  &$\Theta^2_{1,1}(\tau)
      = q^{1/8}(1 - q - 2q^2 + q^3 + 2q^5 + q^6 -2q^9 + \cdots)$
     \tabroom \\

   &$\Theta^2_{2,0}(\tau)
     = q^{9/16}(1 - q - 2q^2 + 2q^4 + 2q^5 + q^7 - 2q^8 + \cdots)$
     \tabroom \\

\hline

4 &$\Theta^4_{0,0}(\tau)
     = q^{1/24}(1 - 2q + 2q^3 - q^5 - 2q^6 + q^7 + 2q^{10} + \cdots)$
     \tabroom \\

  &$\Theta^4_{0,2}(\tau)
     = q^{19/24}(1 - q - q^2 + q^5 + q^6 - q^8 + \cdots) $
     \tabroom \\

  &$\Theta^4_{0,4}(\tau)
     = q^{25/24}(1 - q - 2q^3 + 2q^5 + 2q^7 - 2q^{10} + \cdots) $
     \tabroom \\

  &$\Theta^4_{1,1}(\tau)
     = q^{5/48}\left(1 - q - q^2 + q^3 - q^4 + q^5 - q^6 + q^8 + \cdots \right)$
     \tabroom \\

  &$\Theta^4_{1,3}(\tau)
     = q^{29/48}(1 - 2q^2 - q^3 + q^5 +q^6 + 2q^7 - q^9 + \cdots $
     \tabroom \\

  &$\Theta^4_{2,0}(\tau)
     = q^{3/8}(1 - 2q^2 - q^3 + 2q^5 + 2q^8 - q^9 + \cdots) $
     \tabroom \\

  &$\Theta^4_{2,2}(\tau)
     = q^{1/8}(1 - q - q^3 - q^6 + 2q^7 + 2q^9 + \cdots)$
     \tabroom \\

\hline
\end{tabular}
\end{center}

\centerline{{\bf Table 4.}~{\it Independent Hecke indefinite
modular forms at the level $k=1,2,4$.}}

\vskip .2truein

\section{From Geometry to Conformal Field Theory}

At this point all the ingredients are in place to complete the
proof of the theorems formulated in the introduction. It follows
from Weil's conjecture that the modular forms associated to the
curves $E_i$ have levels that agree with the conductors $N_i =
27,64,144$ determined above. These conductors are divisible by the
bad primes, which gives guidance as to what kind of theta products
$\Theta^k_{\ell,m}(q^a) \Theta^k_{\ell',m'}(q^b)$ should be
considered. The bad primes of $E_6$ are $p=2,3$, which shows that
the form $\Theta^1_{1,1}(q^6)$ is a plausible starting point. This
leaves the proof of the identity in Theorem 1 to all orders, and
the explanation of the twist characters. The proof that the third
relation in (\ref{hw-as-thetas}) holds to all orders of the
$q-$expansion follows from a theorem of Sturm quoted in
\cite{su03}.

A sign difference in two modular forms suggests that they are
related via twists.  For a modular form $f(q)=\sum_n a_nq^n$ and a
Dirichlet character
 \beq \chi:~\mathZ \lra K^{\times} \eeq
 with values in a field $K$, the twisted form $f(q)\otimes \chi$
 is defined as in (\ref{twistedform}). An important class of characters
 is provided by Legendre symbols.
These are defined on rational primes as
 \beq
 \chi_n(p) = \left(\frac{n}{p}\right) =
      \left\{\begin{tabular}{c l} ~1 &$n$ is a square in $\IF_p$ \\
      $-1$ &$n$ is not a square in $\IF_p$ \tabroom \\
            \end{tabular}
\right\}
 \lleq{legendre}
 The conductor of
$\chi_n(\cdot)$ is given by $n$ if $n=1~(\rmmod~4)$ and $4n$ for
$n=2,3~(\rmmod~4)$. For non-prime numbers the generalized Legendre
symbol is defined by using the prime decomposition. Every natural
number $m$ can be decomposed into primes as $m=p_1\cdots p_r$ and
the generalized symbol is defined as \beq \chi_n(m) =
\prod_{i=1}^r \left(\frac{n}{p_i}\right).\eeq Computing the
character $\chi_3$ of Theorems 1 and 2 shows that it produces the
claimed sign changes.

The physical interpretation of the twist characters derives from
the fact that the Legendre character $\chi_n$ determines the
factorization behavior of rational primes $p$ in the number fields
$\mathQ(\sqrt{n})$. The following result holds (see e.g.
\cite{g02}).

{\bf Theorem.}~{\it The splitting behavior of rational primes $p$
into prime ideals $\pfrak_i \subset \cO_{\mathQ(\sqrt{n})}$ in the
ring of algebraic integers $\cO_{\mathQ(\sqrt{n})}$ is determined
by the Legendre symbol $\chi_n$ as follows:}
 \beq \chi_n(p)
= \left\{ \begin{tabular}{r l}
                         1  &if $(p) = \pfrak_1\pfrak_2$ \\
                         $-1$ &if $(p)$ is prime \tabroom \\
                         0    &if $(p)=\pfrak^2$ \tabroom \\
         \end{tabular}
  \right\}.
\eeq

The interpretation of these characters arises from the structure
of the quantum dimensions of the underlying conformal field
theory. The modular behavior of the $N=2$ characters follows from
the modular behavior of the string functions $c^k_{\ell,m}$ and
the theta functions $\theta_{m,k}$: the $N=2$ characters transform
like SU(2)--affine characters in their $\ell$ index and as theta
functions in their $q$ and $s$ index. More precisely, the modular
behavior of these characters is given according to Gepner by \beq
\chi^k_{\ell,q,s}(\tau +1,z,u) = e^{2\pi i
\left(\Delta^k_{\ell,q,s}-c_k\right)} \chi^k_{\ell,q,s}(\tau, z,u)
\eeq for the modular shift $\tau \mapsto \tau +1$, where
$\Delta^k_{\ell,q,s}$ are the conformal weights, which, together
with the U(1)$-$charges, are given by
\begin{eqnarray}
\Delta^k_{\ell,q,s}&=&\frac{\ell(\ell+2)-q^2}{4(k+2)} +\frac{s^2}{8}\\[2ex]
Q^k_{\ell,q,s}&=& \frac{q}{k+2} - \frac{s}{2},
\end{eqnarray}
and $c_k$ is the central charge at level $k$.

For the other generator of $\rmSL(2,\ZZ)$ one finds
 \beq
 \chi^k_{\ell,q,s} \left(-\frac{1}{\tau}, \frac{z}{\tau},
u+\frac{z^2}{2\tau}\right)
 = f'(k) \sum_{\ell',q',s'} S^{\ell \ell'}_{qq',ss'}
 \chi^k_{\ell',q',s'}(\tau,z,u),
 \eeq
 where $f'(k)$ is a constant, and
 \beq S^{\ell \ell'}_{qq',ss'} = e^{\pi i
 \frac{qq'}{k+2}} e^{-\pi i \frac{ss'}{2}} \sin\left(\pi
 \frac{(\ell+1)(\ell'+1)}{k+2}\right).
 \eeq
 Defining the generalized quantum dimensions of the $N=2$ theory
 as
 \beq Q^{\ell m}_{qp,st} = \frac{S^{\ell m}_{qp,st}}{S^{\ell 0}_{q0,s0}}
 \eeq
 it follows that the quantum dimensions $Q^{\ell}_{q,s}:= Q^{\ell
0}_{q0,s0}$ are independent of the charge and spin quantum
numbers, and agree with the SU(2) quantum dimensions $Q_{\ell} =
S_{\ell 0}/S_{00}$, where $S$ is the $S-$matrix of the affine
theory $A_1^{(1)}$ determined by the transformation behavior of
the characters \beq \chi_{\ell} \left(-\frac{1}{\tau},
\frac{u}{\tau}\right) = e^{\pi i ku^2/2} \sum_{\ell'} S_{\ell
\ell'} \chi_{\ell'} (\tau, u) \eeq with  \beq S_{\ell \ell'} =
\sqrt{\frac{2}{k+2}}~~\sin\left(\frac{(\ell+1)(\ell'+1)\pi}{k+2}\right),~~~~~
0\leq \ell, \ell' \leq k. \eeq

These numbers are important because even though they do not
directly provide the scaling behavior of the correlation
functions, they do contain information about the conformal weights
as well as the central charge via the relations
 \beq \frac{1}{L(1)} \sum_{i=1}^k
L\left(\frac{1}{Q_{ij}^2}\right) = \frac{3k}{k+2} - 24
\Delta_j^{(k)} +6j, \lleq{nrt} where $L$ is Rogers' dilogarithm
\beq L(z) = Li_2(z) +{\small \frac{1}{2}} \log(z)~\log(1-z) \eeq
and $Li_2$ is Euler's classical dilogarithm \beq Li_2(z) =
\sum_{n\in \IN} \frac{z^n}{n^2}.\eeq It follows that the quantum
dimensions contain the essential information about the spectrum of
the conformal field theory and Rogers' dilogarithm provides, via
Euler's dilogarithm, the map from the quantum dimensions to the
central charge and the conformal weights. A review of these
results and references to the original literature can be found in
\cite{ak92}.

In the present context the fields of quantum dimensions that
appear at the levels $k=1,2,4$ will be of interest. The result,
\beq
 \begin{tabular}{l| c c c }
 Level $k$   &1  &2  &4 \tabroom \\
 \hline
 QD Field       &$\mathQ$  &$\mathQ(\sqrt{2})$  &$\mathQ(\sqrt{3})$
 \tabroom \\
  \end{tabular},
\eeq will explain the twisting behavior that emerges in Theorem 2.

 \vskip .2truein

\section{From Conformal Field Theory to Geometry}

At this point modular forms have been identified for all Gepner
models at $c=3$. It is therefore of interest to reverse the logic
of providing string theoretic interpretations of geometric
quantities by asking which properties string theoretic modular
forms should have in order to lead to reasonable geometries. Once
a list of properties has been established that determine such
forms one can further address the problem of constructing the
geometry of spacetime from the string theoretic conformal field
theory on the world sheet. These two problems are discussed in
this section. The goal is to identify a well-motivated set of
conformal field theoretic quantities, and to see whether spacetime
can be derived from these objects. Concretely, the question
becomes to what extent some of the considerations presented above
can be turned around, and how powerful the properties described so
far are when one aims to identify 'geometrically induced'
quantities on the world sheet.

\subsection{The constraints}

\underline{I. Conformal field theoretic modular forms}

Given the importance of modular invariance for string theory, it
is natural to test whether modularity holds the key to the
construction of spacetime. Starting from modular forms on the
world sheet, one encounters a problem of riches $-$ even after
having recognized the modular forms associated to the characters
of the partition function as the key ingredients, there are a
great many of them. The focus in this paper is on Hecke indefinite
modular forms as the basic building blocks. Only at level $k=1$ is
there a single such modular form. At level two there are three
independent forms, and at level four there are seven. Combining
these forms leads to a large number of forms, not all of which are
useful for geometric purposes.

\vskip .2truein

\underline{II. Dimension and weight}

The fact that there are a large number of potentially useful
modular forms that arise from a given exactly solvable model is
useful, because the conformal field theoretic models that provide
the building blocks of Gepner and Kazama-Suzuki models should
geometrically be thought of as motives, i.e. pieces of the
variety. The same motive can form a building block for different
varieties, as indicated in \cite{su03,ls04}. This suggests that in
order to make progress some assumptions are needed about the
varieties to be constructed. One such piece of information is the
dimension, and the question arises what property of the modular
forms is related to the dimension of the corresponding geometry.

A statement about the dimension of the target space translates
into a statement about the Hodge type of the variety. If this
Hodge type is $(0,r)$ then it is conjectured that if such Hodge
structures are modular they correspond to modular forms of weight
$(r+1)$ \cite{fm95}. In the case of elliptic curves this is known
via the general proof of the Shimura-Taniyama conjecture
\cite{bcdt01, w95}. Hence the assumption of a one-dimensional
target space translates into the requirement that the modular
forms of interest are of weight two.

\vskip .2truein

\underline{III. Cusp forms}

It appears that forms that are of cusp type are the ones that are
most useful in the context of a geometric interpretation. Hence
the focus should be on such forms, that is forms for which
$a_0=0$. Normalizing these cusp forms means that $a_1=1$. This
criterion can be viewed as an argument to consider the SU(2) theta
functions of the conformal field theory, instead of other forms.
Another argument is based on the fact that characters that enter
the partition functions by themselves are not useful quantities in
the present context because their coefficients count
multiplicities, hence are always positive. This explains why
neither the parafermionic characters
$\eta(\tau)c^k_{\ell,m}(\tau)$, nor the SU(2) characters
$\chi^k_{\ell,m}(\tau)$, nor the N$=$2 characters
$\chi^k_{\ell,q,s}(\tau)$ themselves are useful.

\vskip .2truein

\underline{IV. Hecke eigenforms}

At this point the focus has narrowed to modular cusp forms of
weight 2. Not all such forms are expected to allow a geometric
interpretation. The Hasse-Weil modular form of any variety is
defined via an L-function that is a product of factors defined at
each prime. This suggests to focus on conformal field theoretic
modular forms which admit such a product representation. This
means that the relevant forms should be eigenforms of the Hecke
operators. This condition turns out to be a very useful and strong
restriction for elliptic geometry.

\vskip .2truein

\underline{V. Integral exponents}

The Mellin transform of the Hasse-Weil L-function produces
$q-$expansions which have integral exponents. The basic building
blocks of the string partition function, however, involve
$q-$expansions that have rational exponents. In particular the
string functions, and their associated Hecke indefinite modular
forms, have leading orders that are essentially determined by the
(rational) conformal weights of the associated fields. Aiming at
modular forms of weight two with integral exponents leads to the
consideration of products of the form
 \beq
 \Theta^k_{\ell,m}(a\tau) \Theta^{k'}_{\ell',m'}(b\tau)
 \lleq{products}
 such that
 \beq
 at^k_{\ell,m} + bt^{k'}_{\ell',m'} \in \mathZ,
 \eeq
 where $t^k_{\ell,m}$ and $t^{k'}_{\ell',m'}$ are the
 leading orders of the theta functions $\Theta^k_{\ell,m}$
 and $\Theta^{k'}_{\ell',m'}$.

\vskip .2truein

\underline{VI. Level of the modular form}

Geometric modular forms are usually modular only with respect to
some congruent subgroup of the full modular group $\rmSL(2,\ZZ)$,
determined by the level $N$. It was first observed by Weil, on the
basis of experimental computations, that the modular level $N$ is
determined by the degeneration behavior of the curve at those
primes $p$ for which the reduced variety $X/\mathF_p$ is singular,
i.e. the bad primes \cite{w67}. It was this notion of relating the
geometric conductor to the modular level which helped to make the
Shimura-Taniyama conjecture computationally attractive by making
the relation between elliptic curves and modular forms more
concrete.

A priori there are a great many choices for the coefficients $a$
and $b$ in (\ref{products}), leading to modular forms of the same
weight but different modular level $N$. What is needed is a
constraint for the level $N$ that can be formulated in terms of
the conformal field theory. More precisely, one wants to have a
criterion which prescribes at which primes the curve should have
bad reduction, i.e. provides some of the divisors of $N$. This
leads to the consideration of products of the type
 \beq
 \Theta^k_{\ell,m}(ap\tau) \Theta^{k'}_{\ell',m'}(a'p'\tau),
 \eeq
  where $k,k'$ are the levels, and $p,p'$ are bad primes, or powers
  thereof, and $a,a'$ are integers  that need to be determined.

In the present context a suggestive constraint is to require that
the prime factorization of the quantity $(k+g)$, where $g$ is the
dual Coxeter number, involves only the primes of bad reduction
\beq k+g = \prod_{p~{\rm bad~prime}} p. \eeq In the case of
$A_1^{(1)}$ the dual Coxeter number is $g=2$ and the assumption
means that the bad primes are only those that appear in the
factorization of $k+2$.

\vskip .2truein

\underline{VII. Twisting and quantum dimensions}

At this point enough constraints have been identified to select a
very small set of conformal field theoretic forms. It turns out,
however, that some of them describe the 'wrong' geometry, i.e. not
the geometry of either $E_3,E_4$ or $E_6$. This is not necessarily
unexpected because there is additional information in the
conformal field theory that at this point has not been encoded
geometrically. What is missing is the input of the quantum
dimensions. The correct geometric form emerges only after
multiplying the forms constructed at this point by the Legendre
character associated to the field generated by the quantum
dimensions. This means that for the modular forms $f(q) = \sum_n
a_nq^n$ identified by the conditions formulated above, its twisted
version $f(q) \otimes \chi$ defined in (\ref{twistedform}) should
be considered. In Section 3 it was shown that the fields generated
by the quantum dimensions at the levels $k=1,2,4$ are given by
$\mathQ$, $\mathQ(\sqrt{2})$, and $\mathQ(\sqrt{3})$,
respectively. Associated to these fields are the quadratic
characters $\chi_n$, defined by the Legendre symbols
(\ref{legendre}), and these characters provide the necessary
twists.

\subsection{Applications}

\subsubsection{$(1)^{\otimes 3}$}

It was shown above that at conformal level $k=1$ there is only one
Hecke indefinite modular form $\Theta^1_{1,1}(\tau)$. The
assumptions that the forms should be cusp forms and of weight two
lead to the set of forms of the type
$\Theta^1_{1,1}(a\tau)\Theta^1_{1,1}(b\tau)$ with $a,b \in
\mathN$. The next step is to restrict these cusp forms to those
with integral exponents. This turns out to be a strong
requirement, leading to a set of six forms \beq (a,b) \in
\{(1,11), (2,10), (3,9), (4,8), (5,7), (6,6)\}. \eeq The Hecke
criterion is not very strong at this point, excluding only the
form with $(a,b)=(5,7)$. This leaves five cusp Hecke eigenforms
with integral exponents.

The modular level condition requires that the form should have a
conductor which is divisible only by the prime $k+2=p=3$. This
final constraint uniquely determines the form
$\Theta^1_{1,1}(q^3)\Theta^1_{1,1}(q^9)$. The field of quantum
dimensions at $k=1$ is given by the rational numbers, hence this
form should not be twisted. It follows that the conditions
formulated above uniquely determine the modular form identified in
\cite{su03} as the Mellin transform of the Hasse-Weil L-function
of the cubic Fermat curve. The construction of the geometry is
described below.

\subsubsection{$(2)^{\otimes 2}$}

At conformal level $k=2$ there are three independent Hecke
indefinite modular forms and the resulting number of a priori
possible forms
$\Theta^2_{\ell,m}(a\tau)\Theta^2_{\ell',m'}(b\tau)$ is much
larger than at $k=1$. The requirements of having forms of weight
two, cusp type, and integral exponents leaves a large number of
forms, most of which, however, are not Hecke eigenforms.
Multiplicativity reduces the possible forms to three of type
$\Theta^2_{1,1}(a\tau)\Theta^2_{1,1}(b\tau)$ with \beq (a,b) \in
\{(1,7), (2,6), (4,4)\}. \eeq

The level constraint at $k=2$ implies that the only prime of bad
reduction is $p=2$. This uniquely identifies the form as
$\Theta^2_{1,1}(4\tau)^2$.

The quantum dimension argument finally determines the unique cusp
Hecke eigenform of weight two with integral exponents to be given
by $\Theta^2_{1,1}(4\tau)^2 \otimes \chi_2$, which is the form
determined in \cite{ls04} as the Mellin transform of the quartic
elliptic Brieskorn-Pham curve.

\subsubsection{$(4\otimes 1)$}

For the last of the Gepner models at $c=3$ it is useful to employ
a trick, because the number of a priori possible forms is rather
large. The idea is to use the fact that all Calabi-Yau varieties
are projective and to use inhomogeneous coordinates to analyze the
variety. This leads to the conclusion that the only nontrivial
modular ingredient of the model is the theta function at level
$k=1$. Hence the analysis presented earlier can be applied, except
that in the present case the level constraint must be modified to
require that the modular level should be divisible by the primes
$p=2,3$. This uniquely singles out the form
$\Theta^1_{1,1}(6\tau)^2$. The quantum dimension twist finally
leads to the geometric form $\Theta^1_{1,1}(6\tau)^2\otimes
\chi_3$.

\vskip .2truein

 \underline{\bf Summary}

The results of this section concerning the conformal field
theoretic identification of the target space geometry are
summarized in Table 5.

\begin{center}
\begin{tabular}{c| c c c c c}

Curve  &$N$       &$\bk$          &QD Field
                  &Character
                  &CFT form  \tabroom \\
\hline

$E_3$  &27        &$(1,1,1)$ &$\mathQ$
                  &$\chi_1 = {\bf 1}$
                  &$\Theta^1_{1,1}(q^3)\Theta^1_{1,1}(q^9)$    \tabroom \\

$E_4$  &64        &$(2,2,0)$ &$\mathQ(\sqrt{2})$
                  &$\chi_2(\cdot) = \left(\frac{2}{\cdot}\right)$
                  &$\Theta^2_{1,1}(q^4)^2 \otimes \chi_2$  \tabroom \\

$E_6$  &144        &$(4,1,0)$ &$\mathQ(\sqrt{3})$
                  &$\chi_3(\cdot) = \left(\frac{3}{\cdot}\right)$
                  &$\Theta^1_{1,1}(q^6)^2 \otimes \chi_3$  \tabroom \\

\hline
\end{tabular}
\end{center}

\centerline{{\bf Table 5.}~{\it Modular forms associated to the
curves $E_i$, $i=3,4,6$.}}

\section{Modular Geometry}

In the previous sections a set of constraints was formulated which
 uniquely identify particular Hecke cusp eigenforms in certain
conformal field theories. One can therefore reverse the process
described earlier, and ask what kind of geometry can be
constructed from such modular forms. This question can be asked
for cusp forms of arbitrary weight $w$ and level $N$, i.e. for any
form $f\in S_w(\Gamma_0(N))$, and it is known how to derive a
geometric object from such forms via a construction by Deligne and
others. The obstacle that arises for the general case is that the
geometry associated to cusp forms is given by a Kuga-Sato variety,
which is different from the Calabi-Yau geometry. It is not obvious
what the relation is in general between these Kuga-Sato varieties
and Calabi-Yau spaces, if any. One of the virtues of the present
focus on elliptic curves is not only that the theory of
Eichler-Shimura which applies here is simpler, but that it is
possible to apply some theorems of the arithmetic theory of curves
to see how the abelian varieties associated to cusp forms of
weight 2 are related to the elliptic curve itself.

In essence, the theory of Eichler \cite{e54} and Shimura
\cite{s58} establishes a correspondence between cusp forms of
weight 2 and level $N$ and the cohomology for a particular type of
curve $X_0(N)$, constructed from the modular group $\Gamma_0(N)$
as the compactification
 \beq
 X_0(N) = \overline{\cH/\Gamma_0(N)},
 \eeq
 of the quotient curve $Y_0(N) = \cH/\Gamma_0(N)$. More precisely,
 there exists a map
 \beq
 S_2(\Gamma_0(N)) \lra \rmH^0(X_0(N),\Om^1)
 \eeq
 between cusp forms of weight 2, and level $N$, and the differential forms
 of the modular curve $X_0(N)$, defined as
 \beq
 f ~\mapsto ~ 2\pi i f(z)dz.
 \eeq
 Given an element $f(q) =\sum_n a_nq^n \in S_2(\Gamma_0(N))$ and
its coefficient field $K=\mathQ(\{a_n\})$ Eichler and Shimura
construct an abelian variety $A_f$ of dimension \beq
\rmdim_{\mathC}A_f = [K:\mathQ],\eeq where $[K:\mathQ]$ denotes
the degree of $K$. Roughly, the abelian variety is obtained as a
subvariety, or quotient, of the Jacobian $J(X_0(N))$ of the
modular curve $X_0(N)$ determined by the set $\{f^{\si}\}_{\si}$
of conjugate forms obtained from the set of embeddings
 $\si: K \ra \mathC$ of the field $K$.

If $f$ has rational coefficients then $A_f$ is an elliptic curve,
and given any of the string theoretic modular forms derived in
section 4, it is possible to associate an elliptic geometry to it.
The only remaining question is what happens when one starts from
an elliptic curve $E$, computes its Hasse-Weil L-function,
considers the Mellin transform $f_{\rmHW}(E,q)$ of this
L-function, and then determines from this modular form the
elliptic curve via the Eichler-Shimura construction. It turns out
that the resulting curve is isogenous to the original curve

\beq
 {\thicklines
 \begin{picture}(200,60)
        \put(5,56){$E$}
        \put(30,60){\vector(1,0){60}}
        \put(100,56){$L_{\rmHW}(E,s)$}
        \put(120,45){\vector(0,-1){30}}
        \put(100,0){$f_{\rmHW}(E,q)$}
        \put(90,5){\vector(-1,0){60}}
        \put(0,0){$E_{f_{\rmHW}}$}
        \put(12,15){\vector(0,1){30}}
        \put(-5,30){$\cong$}
 \end{picture}
}. \eeq

It was furthermore shown in a sequence of papers by Eichler,
Shimura, Igusa and Carayol that the Dirichlet series $L(f,s)$ of
the modular form is identical to the Hasse-Weil L-series
$L_{\rmHW}(E,s)$. More precisely one has the following result.

{\bf Theorem.}\cite{c86}~{\it Let $f\in S_2(\Gamma_0(N))$ be a
normalized newform with coefficients in $\mathZ$ and let $E_f$ be
the elliptic curve associated to $f$ via the Eichler-Shimura
construction. Then}
 \beq
 L(E_f,s) = L(f,s)
 \eeq
{\it and $N$ is the conductor of $E_f$.}

This theorem is of importance because it relieves us from an
explicit construction of the elliptic curve $E_f$. Such a
construction can be obtained in a straightforward, if tedious, way
if the generators of the group $\Gamma_0(N)$ are known. The
simplest example is provided by the curve of conductor eleven,
which can be found in \cite{k92}.

\section{Complex multiplication}

It was suggested in \cite{lss04} that exactly solvable Calabi-Yau
varieties are distinguished by the fact that they admit complex
multiplication (CM) in the sense formulated in \cite{lps03}. For
complex dimension one the notion of Calabi-Yau CM coincides with
the usual notion of complex multiplication of elliptic curves, and
it is of interest to ask whether the exactly solvable elliptic
curves discussed here admit complex multiplication in the sense of
Shimura-Taniyama. This question will be addressed in the first
subsection, where the CM type of these curves will be described.
This question can also be turned around, and one can ask whether
the modular forms derived from the Mellin transform of the
Hasse-Weil L-function of elliptic curves with complex
multiplication have special properties. If so, then this opens up
the possibility of using this special property as an additional
criterion to select conformal field theoretical modular forms for
which a geometric interpretation might be attempted. This circle
of ideas will be addressed in the second subsection.

\subsection{Geometric complex multiplication}

The easiest way to see that the elliptic curves $E_i$, $i=3,4,6$
admit complex multiplication is by finding the lattices $\Lambda_i
\subset \mathC$ that define the curves as $E_i =
\mathC/\Lambda_i$. Every elliptic curve can be constructed in
terms of a lattice $\Lambda \subset \mathC$, generated by two
periods $\om_i$ such that $\tau = \om_2/\om_1$ has a positive
imaginary part. The notion of complex multiplication arises from
the observation that for some lattices there are homomorphisms
that are more interesting than the obvious multiplication by an
integer. Denote by $\rmEnd(E)$ the group of endomorphisms of the
elliptic curve \beq \rmEnd(E) = \{z\in \mathC~|~z\Lambda \subset
\Lambda\}. \lleq{endalgebra}

{\bf Definition.} {\it An elliptic curve $E=\mathC/\Lambda$ is
said to admit complex multiplication, or to be a CM curve, if
$\rmEnd(E)\neq \mathZ$.}

It follows from the basic notion of the endomorphism algebra
(\ref{endalgebra}) that both the modulus $\tau$ of the curve, and
the endomorphisms, have to be elements of a quadratic imaginary
field. Define an order $\cO$ of a number field $K$ to be a
subgroup of maximal rank.

{\bf Theorem.}~{\it If the elliptic curve
$E_{\tau}=\mathC/\Lambda(\tau)$ is defined via the lattice
$\Lambda(\tau)= \mathZ+\tau \mathZ$, $\rmIm(\tau)>0$, then either
\hfill \break  1) $\rmEnd(E_{\tau}) = \mathZ$, or \hfill \break 2)
$\mathQ(\tau)$ is a quadratic imaginary field and
$\rmEnd(E_{\tau})$ is isomorphic to an order in $\mathQ(\tau)$.}

As a consequence of this result the notion of CM can be viewed
more conceptually as a map which embeds a CM field $F$ into the
endomorphism algebra tensored with the rational field \beq \theta:
~F \lra~ \rmEnd(E) \otimes \mathQ. \eeq This is the notion of CM
used in \cite{lps03}.

The type of lattices $\Lambda_i$ which lead to the curves $E_i$
can be identified from the affine embedding of $\mathC/\Lambda$,
given by the Weierstrass function $\wp_{\Lambda}(z)$ defined on
the complex plane \bea \mathC/\Lambda ~~&\lra &~~\mathC^2 \nn \\ z
~~&\mapsto &~~(\wp_{\Lambda}(z), \wp^{'}_{\Lambda}(z)), \eea where
$\wp^{'}_{\Lambda}$ denotes the derivative. With
$(x,y)=(\wp_{\Lambda}(z), \wp^{'}_{\Lambda}(z))$ this leads to the
Weierstrass equation \beq y^2 = 4x^3 - g_2(\Lambda)x -
g_3(\Lambda), \eeq where $g_k(\Lambda) = \sum_{\om \in
\Lambda\backslash 0} 1/\om^{2k}$ are the Eisenstein series. It
follows from these series that a lattice that admits the Gaussian
ring of integers as a symmetry must have the general form \beq
E_{\mathQ(\sqrt{-1})}: ~~~y^2 = x^3 + ax, \eeq while a lattice
that admits the Eisenstein integers in $\mathQ(\sqrt{-3})$ as a
symmetry leads to elliptic curves of the form \beq
E_{\mathQ(\sqrt{-3})}:~~~y^2 = x^3 + b. \eeq Comparing these types
of curves with the Weierstrass forms of the Brieskorn-Pham curves
discussed above shows that the curves $E_3$ and $E_6$ admit
complex multiplication by the ring of integers $\cO_K$ for
$K=\mathQ(\sqrt{-3})$, and that $E_4$ admits CM by the Gaussian
ring of integers $\cO_K$ for $K=\mathQ(\sqrt{-1})$.

\subsection{Modular complex multiplication}

In this subsection it will be shown that the modular forms
$f_{\rmHW}(E_i,q)$ $i=3,4,6$ considered above admit complex
multiplication as defined by Ribet \cite{r76} (which can be
motivated by a result concerning the geometry of CM elliptic
curves described in \cite{jps66b}).

{\bf Definition.} \hfill \break {\it A modular cusp form $f(q) =
\sum_n a_nq^n$ is said to admit complex multiplication if there
exists a Dirichlet character $\vphi$ such that} \beq \vphi(p)a_p =
a_p \eeq {\it for all primes $p$ in a set of density 1.}

This means, essentially, that the expansion of a CM modular form
is 'sparse', i.e. many coefficients are zero. This can be made
more precise by describing the vanishing behavior more concretely.
A modular form is CM in the sense of Ribet if the coefficients
$a_p$ vanish at all primes that are inert in some quadratic
extension of the rational field $\mathQ$ \cite{r98}. This is a
formulation that is practical enough for concrete tests. In the
remainder of this subsection the modular forms $f_{\rmHW}(E_i,q)$
are shown to admit complex multiplication in the above sense. The
proof given here combines the arithmetic geometric origin of the
modular forms
 with certain number theoretic results. The first step is to use
the fact that the modular forms under consideration are derived
from the Hasse-Weil L-series of the curve $E_i$. In the previous
section the L-series was explicitly computed for low primes. This
was enough because a result of Sturm implies that the modular form
is determined uniquely by a small number of coefficients. In the
present context it is more useful to determine the coefficients of
this L-series in a more systematic way. This can be achieved by
using the Weierstrass form of the curve.

Suppose that an elliptic curve is given in the form \beq E:~~~y^2
= f(x) \eeq with a polynomial of degree three or four. The
Hasse-Weil L-series \beq L(E,s) = \prod_{p\neq 2} \frac{1}{1 -
\frac{a_p}{p^s} + \frac{p}{p^{2s}}}, \eeq where $a_p= p+1 -
\#(E/\mathF_p)$, can be written as \beq a_p = - \sum_{x~\rmmod~p}
\left( \frac{f(x)}{p} \right). \eeq These sums can be explicitly
computed for concrete polynomials $f(x)$, and allow to
systematically determine the behavior of the coefficients $a_p$ at
various types of primes, in particular their vanishing behavior.

The second step then consists of trying to match the vanishing
behavior found in step one with the prime decomposition of a
number field. This can often be achieved because the factorization
of rational primes in quadratic imaginary fields is known.

The recipe of this proof can be applied to the curves discussed
above. Consider e.g. the curve $E_6$. Denote by
$\left(\frac{x^3-1}{p}\right)$ the Legendre symbol.

{\bf Theorem.}\cite{z00} \hfill \break {\it For the curve
$y^2=x^3-1$ the following holds.}  \beq \sum_{x~\rmmod~p}
\left(\frac{x^3-1}{p}\right) = \left\{\begin{tabular}{l l}
 0,   &$p\equiv 2~\rmmod~3$ \nn \\
 $\pm 2c$, &$p\equiv 1~\rmmod~3$, \tabroom \\
 \end{tabular}
 \right\}, \eeq {\it where $p=c^2+3d^2$.}

 This result determines the vanishing behavior of the form
 $f_{\rmHW}(E_6,q)$. It remains to find a number field for which
 the primes at which the coefficients $a_p$ vanish are precisely
 the inert primes. The natural candidate is the field
 $\mathQ(\mu_3)$, for which one finds the following result,
 e.g. in \cite{c89}.

{\bf Theorem.}~{\it Let $\om=\frac{1}{2}(-1+i\sqrt{3})$ be the
cube root of unity. The Eisenstein primes in the field
$\mathQ(\sqrt{-3})$ are given as follows: \hfill \break
 1) The ordinary primes $p \equiv 2~\rmmod~3$. \hfill \break
 2) The prime $1-\om \in \mathZ[\om]$. \hfill \break
 3) The primes $p \equiv 1~\rmmod~3$ factor as $p=\pi \pi'$,
 where $\pi$ and $\pi'$ are primes in
$\mathZ[\om]$ which are not associates of each other, i.e.
equivalent modulo multiplication by the Eisenstein units $\{\pm
1,\pm \om, \pm \om^2\}$.}

This completes the proof for $f_{\rmHW}(E_6,q)$.

The proofs for the two forms $f_{\rmHW}(E_i,q)$ with $i=3,4$ is
similar. The modular form $f_{\rmHW}(E_3,q)$ also has CM by
$\mathQ(\sqrt{-3})$, while $E_4$ has CM by $\mathQ(\sqrt{-1})$.
The latter can be obtained by combining Gauss's cardinality result
\beq \sum_{x~\rmmod~p} \left(\frac{x^3-x}{p}\right) =
\left\{\begin{tabular}{l l}
 0,   &$p\equiv 3~\rmmod~4$ \nn \\
 $\pm 2a$, &$p\equiv 1~\rmmod~4$, \tabroom \\
 \end{tabular}
 \right\}, \eeq
 {\it with $p=a^2+4b^2$}
with the prime factorization of the Gauss number field (see
\cite{c89}).

{\bf Theorem.}~{\it Let $p$ be a rational prime. Then $p$ factors
in the ring $\mathZ[i]$ of Gauss integers as follows: \hfill
\break
 1) If $p=2$ then $p=-i\pi^2$ where $\pi = 1+i$ is a Gaussian prime. \hfill \break
 2) If $p \equiv 3~\rmmod~4$ then $p$ is inert. \hfill \break
 3) The primes $p \equiv 1~\rmmod~3$ factor as $p=\pi \pi'$, where
 $\pi$ and $\pi'$ are primes in $\mathZ[\om]$ and $\pi$ and $\pi'$ are
 unique up to associates, i.e. equivalent modulo multiplication by the Gaussian units $\{\pm 1,\pm i\}$. }

\section{Arithmetic moonshine}

The main result of the present paper shows that for the simplest
class of exactly solvable Calabi-Yau compactifications it is
possible to prove modular identities which establish a link
between arithmetic geometry and affine Kac-Moody algebras.
Generalized Kac-Moody algebras have played a pivotal role in the
proof \cite{b92} of an old relation between modular functions and
the representation theory of the largest finite simple group, the
 monstrous moonshine conjecture of Conway and Norton \cite{cn79}.
This conjecture was motivated by the original observation of McKay
and Thompson that the first few coefficients of the classical
modular function $j$ suggest a relation between $j$ and traces of
representations of the monster group.

In the present context the starting point is provided by the
arithmetic modular forms of higher weight derived from the
Hasse-Weil L-function which were shown above to arise from
Kac-Moody algebras. It is of interest to ask whether these forms
can be related to some finite simple sporadic group, leading to
what might be called arithmetic moonshine. More concretely, the
question is whether the coefficients of modular forms
$f_{\rmHW}(E_i,q)$ admit an interpretation as traces over
representations of one of the sporadic groups. The purpose of this
section is to briefly indicate how this can be achieved by using
results obtained by Mason for the largest Mathieu group $M_{24}$
\cite{m90}, defined as a subgroup of the permutation group of 24
letters, or by Martin for the Conway group, defined as the
automorphism group of the Leech lattice \cite{m96}. A different
application of modular functions in the context of enumerative
aspects of mirror symmetry has been discovered by Lian and Yau
\cite{ly96}, and further investigated by Verrill and Yui
\cite{vy01} and Doran \cite{d01}.

Mason's idea is to associate traces to pairs of elements $(g,h)
\in G\times G$, for any finite group $G$. More precisely, assume
that the pair $(g,h)$ is a rational commuting pair, i.e. the
action of the element $h$ is assumed to be rational on the
eigenspaces of the element $g$. This notion was originally
introduced by Norton in the context of the Fischer-Griess monster.
The basic concept is that of an elliptic system associated to
commuting rational pairs $(g,h)$, defined as a map which
associates to $h$ a graded infinite dimensional complex vector
space $V_h=\oplus_n V_{h,n}q^n$ such that every homogeneous
component of $V_h$ affords a finite dimensional complex
representation for the centralizer of $h$ in $G$, and if $g\in G$
commutes with $h$, then its graded trace $f(g,h;q)=\rmtr_{V_h}(g)$
in $V_h$ is a modular function or modular form. In \cite{m90}
Mason considers the elliptic system of $M_{24}$, and using his
results, extended to the Conway group in \cite{m96}, leads to the
interpretation of the modular forms of the three elliptic
Brieskorn-Pham curves as forms associated to pairs of group
elements $(g,h) \in M_{24}\times M_{24}$
 \beq
 f_{\rm HW}(E,q) = \rmtr_{V_h}(g).
 \eeq
 These relations are summarized in Table 6, where the group
 elements are written in terms of the conjugacy classes of the
 groups. The numerical prefix indicates the order of the elements,
 and the letter symbol is used to distinguish between classes of
 the same order \cite{atlas}. An enumeration of these classes and
 Frame shapes can be found in \cite{m90} and \cite{l89}.

\beq
\begin{tabular}{c| c  c}

Curve $E_i$  &Theta form   &Mathieu pair $(g,h)$  \tabroom \\

\hline

$E_3$  &$\Theta^1_{1,1}(q^3)\Theta^1_{1,1}(q^9)$
       &$(3B,3A)$   \tabroom \\

$E_4$ &$\Theta^2_{1,1}(q^4)^2\otimes \chi_2$
       &$(4A,4A)$   \tabroom \\

$E_6$ &$\Theta^1_{1,1}(q^6)^2\otimes \chi_3$
      &$(6B,2B)$    \tabroom \\

\hline
\end{tabular}
\eeq

\centerline{{\bf Table 6.}~{\it Mathieu classes for
$f_{\rmHW}(E_i,q)$, $i=3,4,6$.}}

The interpretation of the modular forms of elliptic Brieskorn-Pham
curves as Kac-Moody theoretic objects, and as series associated to
sporadic finite simple groups, provides a link between affine Lie
algebras and moonshine that applies to standard string theory
models, without the need of the constructions of \cite{b92}.
Perhaps the notion of arithmetic moonshine will help to gain a
more conceptual understanding of the link between Kac-Moody
algebras and arithmetic geometry described here and in
\cite{su03,ls04}.

\section{Final comment}

The analysis of this paper may appear to have succeeded only in
recovering part of the geometry of spacetime, namely the
arithmetic part that is obtained by reducing the complex variety
that is normally considered. The results obtained so far indicate
that it may well be that this is all that is needed in the general
case to understand the physics of the string from the geometry of
the variety. Put differently, it is possible that all the string
needs are the arithmetic properties of spacetime.

In the present simple case it turns out that the discrete
structure is enough to recover the Calabi-Yau geometry from the
arithmetic properties. The reason for this is a conjecture of Tate
from the 1960s to the effect that abelian varieties are determined
up to isogeny by the Tate module, i.e. essentially by the torsion
points of the abelian variety. This translates into the fact that
two abelian varieties defined over the rational numbers must be
isogenous if and only if their L-functions coincide. It follows
that an equivalence of the arithmetic of two varieties over all
finite fields implies an isogeny over the rationals, dense
everywhere in the reals.

\vskip .3truein

{\bf ACKNOWLEDGEMENT.}

It is a pleasure to thank Monika Lynker for conversations and Jack
Morse for correspondence. Part of this work was completed while
the author was a Scholar at the Kavli Institute for Theoretical
Physics in Santa Barbara. This work was supported in part by the
National Science Foundation under Grant No. PHY99$-$07949 and a
KSU Incentive Grant for Scholarship. Most of the results of this
work have been presented in talks at the Schr\"odinger Institute
in May 2004 and the Perimeter Institute in December 2004. It is a
pleasure to thank the KITP, the Schr\"odinger Institute, and the
Perimeter Institute for hospitality.

\end{document}